\documentclass[12pt]{article}

\usepackage{amssymb,amsfonts,amsthm}
\usepackage{amsmath}
\usepackage{bm}             
\usepackage[mathscr]{eucal}
\usepackage{cancel}
\usepackage{wasysym}

\def\be{\begin{equation}}
\def\ee{\end{equation}}
\def\bea{\begin{eqnarray}}
\def\eea{\end{eqnarray}}

\def\bdelta{\mbox{\boldmath $\delta$}}

\def\hsp5{\hspace{5mm}}

\theoremstyle{remark}

\newcommand{\sfrac}[2]{{\textstyle{#1\over#2}}}

\title{\sc Second order cosmological perturbations: simplified gauge change
formulas}

\begin{document}

\author{ \\
{\Large\sc Claes Uggla}\thanks{Electronic address:
{\tt claes.uggla@kau.se}} \\[1ex]
Department of Physics, \\
Karlstad University, S-651 88 Karlstad, Sweden
\and \\
{\Large\sc John Wainwright}\thanks{Electronic address:
{\tt jwainwri@uwaterloo.ca}} \\[1ex]
Department of Applied Mathematics, \\
University of Waterloo,Waterloo, ON, N2L 3G1, Canada \\[2ex] }


\date{}
\maketitle

\begin{abstract}

In this paper we present a new formulation of the change of gauge formulas
in second order cosmological perturbation theory which unifies and simplifies
known results. Our approach is based on defining new second
order scalar perturbation variables by adding a
multiple of the square of the corresponding first
order variables to each second order variable. A bonus is that these new perturbation
variables are of broader significance in that they also simplify the analysis of second
order scalar perturbations in the super-horizon regime in a number of ways, and lead to new
conserved quantities.

\end{abstract}

\section{Introduction}

Cosmological perturbation theory plays a central role in confronting theories
of the early universe with observations. The increasing accuracy of the observations,
however, has made it desirable to extend the theory
from linear to second order (\emph{i.e.\,}nonlinear)
perturbations,\footnote
{See, for example, Bartolo \emph{et al} (2010)~\cite{baretal10} and
Tram \emph{et al} (2016)~\cite{traetal16}.}
 which presents various technical challenges.
For example, in applying cosmological perturbation theory at second
order it is often desirable to use several
gauges since the physical interpretation may require
one gauge while the mathematical analysis
may be simpler using a different gauge. The change
of gauge formulas at second order that are
needed in this situation have a complicated
structure and are cumbersome to work with
due to the presence of source terms that
depend quadratically on the first order perturbations.
This state of affairs motivated us to revisit the problem of
gauge change at second order. Our starting point
was two papers that give a fairly comprehensive
treatment of this topic, with a description of the
general method for deriving the formulas,
together with various specific cases, namely,
Noh and Hwang (2004)~\cite{nohhwa04}
(see sections VI and VII) and Malik and Wands (2009)~\cite{malwan09}
(see sections 6 and 7).\footnote{Other papers that
have been influential in developing and
applying second order cosmological perturbation
theory but do not emphasize change of gauge
formulas are Bartolo \emph{et al} (2004)~\cite{baretal04a}
and Nakamura (2007)~\cite{nak07}.
Examples of recent papers that use change of gauge formulas at second order are
Malik (2005)~\cite{mal05}, Christopherson \emph{et al} (2011)~\cite{chretal11b}, Hidalgo
\emph{et al} (2013)~\cite{hidetal13}, Christopherson \emph{et al} (2015)~\cite{chretal15},~Carrilho
and Malik (2015)~\cite{carmal15}, Dias \emph{et al} (2015)~\cite{diaetal15}, Villa and
Rampf (2016)~\cite{vilram16} (see equations (3.11)-(3.14) and (3.22)-(3.25) with the source
terms given in equations (C.1-C.4)) and Hwang \emph{et al} (2017)~\cite{hwaetal17}.}
Reading and interpreting the formulas in these papers is
not easy due to their complexity, and comparing formulas in the papers is difficult
because of the lack of a standard notation. However, while studying the formulas in these
papers we noticed that for scalar perturbations
they have certain features in common that enables
one to write them in a unified and simpler form.

We consider first and second order scalar perturbations of
Friedmann-Lema\^{i}tre (FL)
universes subject to the following assumptions:
\begin{itemize}
\item[i)] the spatial background is flat;
\item[ii)] the stress-energy tensor can be written in the form
$T^a\!_b = \left(\rho + p\right)\!u^a u_b + p\delta^a\!_b$,
thereby describing perfect fluids and scalar fields;
\item[iii)] the linear perturbation is purely scalar.
\end{itemize}
We will use  the following five gauges: the Poisson
(longitudinal, zero shear) gauge, the uniform
curvature (spatially flat) gauge, the total matter gauge,\footnote{We refer to
Malik and Wands (2009)~\cite{malwan09}, section 7.5,
for this terminology. See also Liddle and Lyth (2000)~\cite{lidlyt00}, page 343.  This
gauge was apparently introduced
by Kodama and Sasaki (1984)~\cite{kodsas84},
and called the velocity-orthogonal isotropic gauge (see page 45, case 2b).}
the uniform density gauge, and the uniform scalar field gauge.

The simplification of the gauge change formulas is
accomplished by making three choices.
 First, we use a common
fixing of the spatial gauge freedom so that the remaining degrees of
freedom in the gauge vector fields ${}^{(r)}\!\xi^a,\,r=1,2$, are the
temporal components at first and second order.
Second, we normalize background and perturbation variables
so that they are dimensionless.
In particular, as the time variable we
use the so-called \emph{$e$-fold time $N$}
which is defined by $N = \ln(a/a_0)$, where $a$
is the background scale factor.\footnote
{See for example, Martin and Ringeval (2006)~\cite{marrin06}. }
The scalar $N$ represents the number of
background $e$-foldings from some reference time $a_0$.
Third, a careful inspection
of the source expressions in the known gauge
change formulas reveals that a number of quadratic
first order terms can be incorporated in a systematic way into the second order
perturbation variables and the temporal gauge vector field,
leaving much simplified source terms. We will use a hat notation
${\hat f}$ for these \emph{source-compensated  second order
perturbation variables}. For the metric and matter variables
we can give a unified definition as follows:
\begin{equation}
{}^{(2)}\!{\hat{\Box}} :=  {}^{(2)}\!{\Box} + C_{\Box} {}^{(1)}\!{\Box}^2, \label{hat_box1}
\end{equation}
where the kernel $\Box$ represents a dimensionless metric or matter
perturbation and the coefficient $C_{\Box}$ depends on the
background variables, while for the temporal component
of the gauge vector field using $e$-fold time $N$ as time coordinate we define
\begin{equation}
{}^{(2)}\!{\hat{\xi}}^N := {}^{(2)}\!{\xi}^N -
{}^{(1)}\!{\xi}^N \partial_{N} ({}^{(1)}\!{\xi}^N), \label{hat_z1}
\end{equation}
where we write $\partial_N\equiv\partial/\partial_N$ for brevity.
In terms of these quantities,
inspection of the known change of gauge formulas
leads to the following unified form:
\begin{subequations}\label{gaugetransf}
\begin{align}
{}^{(1)}\!{\Box}_{\bullet} &= {}^{(1)}\!{\Box} -
{}^{(1)}\!\xi^N_{\bullet}, \label{trans_gen1}\\
{}^{(2)}\!{\hat{\Box}}_{\bullet} &=  {}^{(2)}\!{\hat{\Box}} -
{}^{(2)}\!\hat{\xi}^N_{\bullet} +
2{}^{(1)}\!\xi^N_{\bullet}\, \partial_{N}\! {}^{(1)}\!{\Box}_{\bullet}  +
\Box_{rem,\bullet}, \label{trans_gen2}
\end{align}
\end{subequations}
where the subscript $_\bullet$ stands for a letter describing a particular gauge choice.
We find that the reminder term $\Box_{rem,\bullet}$ for most metric and matter variables
is a simple quadratic function of the first order variables, which in the case of any scalar
variable is in fact zero.

The outline of the paper is as follows. In section~\ref{unified},
after introducing the notation that we
will use for the metric and matter variables, we present
the details concerning the unified
definition~\eqref{hat_box1} of the hat variables and the
details concerning the unified form of the
gauge transformation formula~\eqref{gaugetransf}. In
section~\ref{change_of_gauge} for each of the five
choices of temporal gauge indicated by $_\bullet$ we give
expressions for ${}^{(2)}\!{\hat{\Box}}_{\bullet}$
in terms of the metric and matter perturbation variables
(see equations~\eqref{box_change2}).
This set of formulas, which provides an efficient unifying algorithm for
calculating any gauge invariant in any of the
above five gauges, is the main goal of the paper. In
section~\ref{applications} we use our unified scheme to
give simple derivations of some important change of gauge formulas previously
presented in the cosmological literature.
In section~\ref{discussion} we point out that the present paper is the first
of four closely connected papers. We also comment on how the new hatted
variables result in new conserved quantities, as shown in detail in
the sequel papers. In appendix~\ref{comp_lit} we make further
comparisons of our gauge transformation formulas in
section~\ref{unified} with those in Malik and
Wands (2009)~\cite{malwan09}, which served as our main
starting point for the present paper.

\section{Unified form for gauge transformations to second order \label{unified}}

To perturb a flat FL background geometry it is convenient to write the metric as
\begin{equation} \label{pert_metric}
ds^2 = a^2\left(-(1+2\phi)d\eta^2 +  f_{\eta i}\,d\eta dx^i + f_{ij}dx^i dx^j \right),
\end{equation}
where $a$ is the background scale factor and $\eta$ is
conformal time in the background.
We assume that the metric components can be expanded
in powers of a perturbation parameter
$\epsilon$, \emph{i.e.} as a Taylor series, for example,
\begin{equation}
\phi = \epsilon\,{}^{(1)}\!\phi + \sfrac12 \epsilon^2\,{}^{(2)}\!\phi + \dots\, .
\end{equation}
We furthermore assume that the metric can be decomposed
into scalar, vector, and tensor perturbations
according to
\begin{subequations}\label{metric}
\begin{align}
f_{\eta i} &=  {\bf D}_i B + B_i, \\
f_{ij} &= (1-2\psi)\gamma_{ij}+ 2{\bf D}_i {\bf D}_j C + 2{\bf D}_{(i }C_{j)} + 2C_{ij},
\end{align}
\end{subequations}
where ${\bf D}^iB_i=0; {\bf D}^iC_i=0$; $C^i\!_i=0$, ${\bf D}^iC_{ij}=0$, and
where ${\bf D}_i$ is the spatial
covariant derivative corresponding to the flat
metric $\gamma_{ij}$. Use of Cartesian background coordinates yields
$\gamma_{ij} = \delta_{ij}$ and ${\bf D}_i = \partial/\partial x^i$.
As regards dimensions, we make the choice that the scale factor $a$ is dimensionless.
It then follows that the coordinates $\eta$ and $x^i$ have dimensions of \emph{length}
since $ds^2$ has dimension $\text{\emph{length}}^2$.
Hence $\phi$ and $\psi$ are dimensionless while $B$ has
dimension \emph{length}.

We consider a stress-energy tensor of the form:
\begin{equation}\label{pf}
T^a\!_b = \left(\rho + p\right)\!u^a u_b + p\delta^a\!_b ,
\end{equation}
which encompasses perfect fluids and scalar fields.
The energy density $\rho$, the pressure $p$, and the
4-velocity $u^a$ can be expanded as a Taylor series in $\epsilon$.
Perturbations of the energy density $\rho$ are therefore given by\footnote
{We use a subscript zero to denote the background value of some quantity,
so that $\rho_0$ and $p_0$ are the background energy density and pressure.}
\begin{equation}
\rho=\rho_0 + \epsilon\,{}^{(1)}\!\rho +
\sfrac12 \epsilon^2\,{}^{(2)}\!\rho + \dots,
\end{equation}
and similarly for the pressure perturbations. We use the usual
background matter variables $w$
and $c_s^2$, and the deceleration parameter $q$ defined according to
\begin{equation}
w = \frac{p_0}{\rho_0}, \qquad  c_s^2 = \frac{p_0'}{\rho_0'},
\qquad q= -\frac{{\cal H}'}{{\cal H}^2},   \label{w,c_s}
\end{equation}
where  ${}^\prime$ denotes the derivative with respect to the conformal background
time variable $\eta$, and ${\cal H}=a'/a=aH$ with $H$ the background Hubble variable.
We use units such that $c=1$ and $8\pi G=1$, where
$c$ is the speed of light and $G$ the gravitational constant.
It follows that ${\cal H}$ has dimension of $(\emph {length})^{-1}$ and that $q, w$ and $c_s^2$ are dimensionless.

Since we have assumed that the spatial background is flat, the Einstein field
equations in the background can be written as
\begin{equation}
3{\cal H}^2 = a^2\rho_0, \qquad 2(-{\cal H}'+{\cal H}^2)  =a^2(\rho_0 + p_0),
\end{equation}
which in conjunction with~\eqref{w,c_s} yields the following relation
between $w$ and the deceleration parameter $q$:
\begin{equation} \label{q,w.relation}
1 + q = \sfrac32(1+w),
\end{equation}
a result that we will use frequently.

To define the scalar velocity perturbations we find it convenient to
work with the \emph{covariant} 4-velocity $u_b$, which we normalize
with a conformal factor $a$ according to $u_b=a V_b$, in analogy
with the conformal factor $a^2$ in the metric~\eqref{pert_metric}.
We then expand and decompose the
spatial components of $V_b$ according to
\begin{subequations} \label{vel_exp}
\begin{align}
V_i &= \epsilon\,{}^{(1)}\!{V}_i +
\sfrac12 \epsilon^2\,{}^{(2)}\!{V}_i + \dots, \label{vel_exp1} \\
{}^{(r)}\!{V}_i &= {\bf D}_j {}^{(r)}\!{V} + {}^{(r)}\!{\tilde V}_i,\qquad r = 1, 2,\dots ,
\end{align}
\end{subequations}
with ${\bf D}^i{}^{(r)}\!{\tilde V}_i=0$, so that ${}^{(r)}\!{V}$
represents the scalar perturbations.
Since the $V_b$ are dimensionless and the $x^i$ have dimension
\emph{length} it follows from~\eqref{vel_exp}
that ${}^{(r)}\!V$ has dimension \emph{length}.

In the case in which the
matter-energy content is provided by a minimally-coupled
scalar field, we will use $\varphi$ to denote the scalar field and define the perturbations according to
\begin{equation}
\varphi = \varphi_0 + \epsilon {}^{(1)}\!{\varphi} +
\sfrac12 \epsilon ^2\,{}^{(2)}\!{\varphi} + \dots,.
\end{equation}

Next we turn to gauge transformations in cosmological perturbation theory.
We begin by considering an arbitrary 1-parameter family of a tensor field $A(\epsilon)$,
which can be expanded in powers of $\epsilon$, {\it i.e.} as a Taylor
series:
\begin{equation}\label{taylor} \mathrm{A}(\epsilon) = \mathrm{A}_0 +
\epsilon\,{}^{(1)}\!\mathrm{A} + \sfrac{1}{2}\epsilon^2\,
{}^{(2)}\!\mathrm{A} + \dots\, .
\end{equation}
A gauge transformation induces a change in the first and second order perturbations of
$A(\epsilon)$. Arguably the most geometric and straightforward approach to gauge transformations
is the ``active approach'' using an exponential map described in section 6 in Malik and Wands
(2009)~\cite{malwan09}, and this is the approach we take as our starting point.\footnote{Gauge
transformations up to second order in cosmological
perturbation theory can also be represented in coordinates as follows
(see e.g. Malik and Wands (2009)~\cite{malwan09}):
$$ {\tilde x}^a = x^a +\epsilon {}^{(1)}\!\xi^a + \sfrac12
\epsilon^2\left({}^{(2)}\!\xi^a+
{}^{(1)}\!\xi^a\!_{,b}{}^{(1)}\!\xi^b\right).$$}
First and second order gauge transformations are then represented as
(equations (6.5) and (6.6), respectively, in~\cite{malwan09}):
\begin{subequations}\label{delta_A_begin}
\begin{align}
{}^{(1)}\!{A}[\xi] &= {}^{(1)}\!A +
\pounds_{{}^{(1)}\!\xi}A_0, \\
{}^{(2)}\!{A}[\xi] &= {}^{(2)}\!A + \pounds_{{}^{(2)}\!\xi}A_0 +
\pounds_{{}^{(1)}\!\xi}\left(2{}^{(1)}\!A +
\pounds_{{}^{(1)}\!\xi}A_0\right),
\end{align}
\end{subequations}
where ${}^{(1)}\!\xi^a$ and ${}^{(2)}\!\xi^a$ are independent
background gauge vector fields and $\pounds$ is the Lie derivative
(see also~\cite{bruetal97}, equations (1.1)--(1.3)).
Equations~\eqref{delta_A_begin} describe how the
tensor field $A$ changes under an arbitrary
gauge transformation. More importantly from a
physical point of view, these equations serve to define
gauge invariant quantities in the following way. If we
impose a restriction on the perturbation
variables that determines the gauge fields uniquely,
say ${}^{(1)}\!\xi^a = {}^{(1)}\!\xi_{\bullet}^a$, ${}^{(2)}\!\xi^a = {}^{(2)}\!\xi_{\bullet}^a$,
then we say that we have fixed the gauge. If we
use these as the gauge fields in~\eqref{delta_A_begin}, then the
quantities ${}^{(1)}\!{A}[\xi_{\bullet}]$ and ${}^{(2)}\!{A}[\xi_{\bullet}]$
so defined are gauge invariant quantities. On introducing the
shorthand notation
\begin{equation} \label{A_bullet}
{}^{(1)}\!{A}_{\bullet} = {}^{(1)}\!{A}[\xi_{\bullet}], \qquad
{}^{(2)}\!{A}_{\bullet} = {}^{(2)}\!{A}[\xi_{\bullet}],
\end{equation}
equations~\eqref{delta_A_begin} yield
\begin{subequations}\label{delta_A}
\begin{align}
{}^{(1)}\!{A}_{\bullet} &= {}^{(1)}\!A +
\pounds_{{}^{(1)}\!\xi_{\bullet}}A_0,  \label{delta_A1}\\
{}^{(2)}\!{A}_{\bullet} &= {}^{(2)}\!A + \pounds_{{}^{(2)}\!\xi_{\bullet}}A_0 +
\pounds_{{}^{(1)}\!\xi_{\bullet}}\left(2{}^{(1)}\!A +
\pounds_{{}^{(1)}\!\xi_{\bullet}}A_0\right).
\label{delta_A2}
\end{align}
\end{subequations}
We say that ${}^{(1)}\!{A}_{\bullet}$ and  ${}^{(2)}\!{A}_{\bullet}$
are the first and second order gauge invariants associated with
the tensor field $A$ in the gauge specified by the subscript ${}_{\bullet}$.
We list several gauges and their identifying subscripts at the
beginning of section~\ref{change_of_gauge}.

We \emph{fix the spatial gauge} freedom completely by setting the metric functions $C$ and
$C_i$ in~\eqref{metric} to be zero order by order, which up to second order
gives
\begin{equation}\label{Ccond}
{}^{(r)}\!C = 0, \qquad {}^{(r)}\!C_i = 0,\qquad r=1,2.
\end{equation}
The above spatial gauge fixing is arguably the essence of Hwang and Noh's so-called
``gauge ready'' approach, who refer to it as the C-gauge (see for example
Noh and Hwang (2004)~\cite{nohhwa04}, equation (259)).
Note that this is the only way one can \emph{algebraically}
completely fix the spatial gauge by
using the metric components and matter variables
for the present models (see e.g. the gauge
transformations given in~\cite{uggwai14a}), and as a consequence all the gauges listed in
the introduction and section~\ref{change_of_gauge} are characterized by this condition.
The only gauge that is commonly used that does not
fulfil this condition is the synchronous gauge, which is useful for treating
dust models.\footnote{For a recent work using the synchronous gauge for models with dust,
see e.g. Gressel and Bruni (2017)~\cite{grebru17}.} However, the synchronous gauge
is not a fully fixed gauge and the natural way to completely fixing this gauge,
and thereby relate quantities to physical observables, is to relate it to
the total matter gauge, which does obey the above conditions, see
Appendix B.7 in~\cite{uggwai14a}.

As a consequence of the above spatial gauge fixing, the remaining gauge freedom
is described by gauge fields to second order restricted to be of the form
\begin{equation}
{}^{(1)}\!\xi^a = ({}^{(1)}\!\xi^N, 0), \qquad
{}^{(2)}\!\xi^a = ({}^{(2)}\!\xi^N, {\bf D}^i {}^{(2)}\!\xi + {}^{(2)}\!{\tilde \xi}^i ), \qquad
{\bf D}_i {}^{(2)}\!{\tilde \xi}^i=0,
\label{xi_restricted}
\end{equation}
where ${}^{(2)}\!\xi$ and ${}^{(2)}\!{\tilde \xi}^i$
are determined by quadratic source terms that
arise from the conditions~\eqref{Ccond}, where,
in particular, ${}^{(2)}\!\xi$ depends on
${}^{(1)}\!\xi^N$.\footnote{See equation (B10e) and (B10f)
in~\cite{uggwai14a} for the transformation
laws for $C$ and $C_i$.} As in the introduction we are using the $e$-fold
 time defined by $N =\ln(a/a_0)$ as the time
coordinate instead  of the conformal time $\eta$.
Note that the temporal components of the
gauge fields are related according to $\xi^N = {\cal H}\xi^\eta$,
which follows from
\begin{equation}
\partial_\eta = {\cal H}\partial_N. \label{eta_tau}
\end{equation}
In this paper we will primarily use $e$-fold time $N$ but we
will also use conformal time $\eta$, depending on the context.
Equation~\eqref{eta_tau} enables one to
make the transition and we will use it frequently.

We are further restricting our considerations to perturbations
that are \emph{purely scalar at linear order}, \emph{i.e.} the metric functions
that describe vector and tensor perturbations at first order are zero:
\begin{equation} \label{pure.scalar}
{}^{(1)}\!B_i = 0, \qquad {}^{(1)}\!C_{ij} = 0, \qquad {}^{(1)}\!{\tilde V}_i = 0.
\end{equation}
Bartolo \emph{et al} (2004)~\cite{baretal04a} (see page 41) argue
that this restriction is reasonable on physical grounds,
since vector perturbations have decreasing amplitude and are not
generated during inflation, while tensor perturbations
are expected to be negligible.
On the other hand it is well known (\cite{baretal04a}, see page 41)
that even if the vector and tensor
perturbations are zero at first order, they will be generated
at second order due to the presence of source terms in the vector and tensor
governing equations, since these source terms depend on the first order scalar
perturbations. Thus even if the first order restriction~\eqref{pure.scalar}
holds we will have
\begin{equation}
{}^{(2)}\!B_i \neq 0, \qquad {}^{(2)}\!C_{ij} \neq 0, \qquad {}^{(2)}\!{\tilde V}_i \neq 0,
\end{equation}
at second order.
In this context, however, the second order scalar perturbations
are independent of the second order vector and tensor perturbations
and hence can be studied separately.
In this paper we are choosing to consider
only the scalar perturbations at second order,
which physically represent density perturbations,
leaving the second order vector and tensor perturbations
for future work.\footnote{The tensor mode at
second order describes gravitational waves generated by the first order
scalar (\emph {i.e.} matter) perturbations.}
We are thus studying second order scalar perturbations
subject to the first order restriction~\eqref{pure.scalar},
and they
are represented by the functions ${}^{(r)}\!\phi$, ${}^{(r)}\!B$, ${}^{(r)}\!\psi$,
${}^{(r)}\!V$, ${}^{(r)}\!{\rho}$, ${}^{(r)}\!p$, ${}^{(r)}\!\varphi$,
and the remaining gauge freedom which is described by the functions ${}^{(r)}\!\xi^N$, $r=1,2$.

We now describe how the first order variables
${}^{(1)}\!B, {}^{(1)}\!\psi, {}^{(1)}\!V, {}^{(1)}\!{\rho},{}^{(1)}\!\varphi$
transform under the remaining temporal gauge
freedom. From~\eqref{delta_A1} one obtains the well-known relations:
\begin{subequations}
\begin{xalignat}{2}
{}^{(1)}\!B_{\bullet} &= {}^{(1)}\!B -{\cal H}^{-1} {}^{(1)}\!\xi^N_{\bullet}, &\qquad
{}^{(1)}\!\psi_{\bullet} &= {}^{(1)}\!\psi - {}^{(1)}\!\xi^N_{\bullet}, \\
{}^{(1)}\!V_{\bullet} &= {}^{(1)}\!V -{\cal H}^{-1} {}^{(1)}\!\xi^N_{\bullet}, &\qquad
{}^{(1)}\!A_{\bullet} &= {}^{(1)}\!A  + (\partial_N{A}_0)({}^{(1)}\!\xi^N_{\bullet}),
\end{xalignat}
\end{subequations}
where $A=\rho$ or $A=\varphi$. By normalizing the perturbations (apart
from $\psi$) these transformation rules can be written in the unified form
given in Eq.~\eqref{trans_gen1}, which we repeat here:
\begin{equation}{}^{(1)}\!{\Box}_{\bullet} = {}^{(1)}\!{\Box} -
{}^{(1)}\!\xi^N_{\bullet}, \label{trans_gen1_repeat}
\end{equation}
where the kernel $\Box$ represents the following
variables in the five different cases:
\begin{equation}
\Box = \psi, \quad \Box = {\cal H}B, \quad  \Box  = {\cal H}V,
\quad \Box =\frac{\rho}{(-\partial_N{\rho}_0)}, \quad
\Box = \frac{\varphi}{(-\partial_N{\varphi}_0)}.  \label{box_var}
\end{equation}
The above normalization ensures that the variables
are dimensionless (recall that $B$ and $V$ have dimension \emph{length}
and $\cal H$ has dimension $(\emph{length})^{-1}$).

We now consider the second order perturbation variables.
The second order gauge transformation formulas,
which follow from~\eqref{delta_A2}, can
be written so as to have the same leading
order terms as the first order
formula~\eqref{trans_gen1_repeat} but they also include
a source term ${\mathbb S}_{\Box}$ that depends quadratically on the
first order variables in a sometimes complicated way:
\begin{equation}\label{box2}
{}^{(2)}\!\Box_{\bullet}= {}^{(2)}\!\Box - {}^{(2)}\!\xi^N_{\bullet} + {\mathbb S}_{\Box}.
\end{equation}
As described in the introduction, inspection of the different expressions
${\mathbb S}_{\Box}$ reveals that a number of quadratic first
order terms can be incorporated
into the second order terms in the formula~\eqref{box2}, leaving
much simplified source terms. The
resulting unified formula is given by equation~\eqref{trans_gen2},
which we repeat here for the reader's convenience:
\begin{equation}
{}^{(2)}\!{\hat{\Box}}_{\bullet} =  {}^{(2)}\!{\hat{\Box}} - {}^{(2)}\!\hat{\xi}^N_{\bullet} +
2{}{}^{(1)}\!\xi^N_{\bullet}\, \partial_{N}\! {}^{(1)}\!{\Box}_{\bullet}  +
\Box_{rem,\bullet}, \label{trans_gen2_repeat}
\end{equation}
where $\Box$ stands for one of the variables in~\eqref{box_var}.
The hatted variables are given by
equations~\eqref{hat_box1} and~\eqref{hat_z1}, which we repeat here:
\begin{subequations}
\begin{align}
{}^{(2)}\!{\hat{\Box}} &=  {}^{(2)}\!{\Box} + C_{\Box} {}^{(1)}\!{\Box}^2, \label{hat_box} \\
{}^{(2)}\!{\hat{\xi}}^N_{\bullet} &= {}^{(2)}\!{\xi}^N_{\bullet} -
{}^{(1)}\!{\xi}^N_{\bullet}\, \partial_{N} ({}^{(1)}\!{\xi}^N_{\bullet}). \label{hat_z}
\end{align}
\end{subequations}
The details of this unified formulation lie in the coefficients
$C_{\Box} $ in~\eqref{hat_box} and
in the remaining source terms $\Box_{rem,\bullet}$ in~\eqref{trans_gen2_repeat}.
The expressions for these quantities are obtained by
comparing the transformation rules obtained
from~\eqref{delta_A2} for each choice of $\Box$ in~\eqref{box_var} with the unified
form~\eqref{trans_gen2_repeat}.

First, the coefficients $C_{\Box} $ in~\eqref{hat_box} are given by:
\begin{equation} \label{C_box}
C_{\psi} = 2, \qquad  C_B = C_V = 1+q, \qquad
C_{\rho} = \frac{\partial^2_{N}\rho_0}{\partial_{N}\rho_0}, \qquad
C_{\varphi} = \frac{\partial^2_{N}{\varphi_0}}{\partial_{N}\varphi_0}.
\end{equation}
For a non-interacting fluid or a non-interacting scalar
field,\footnote{The stress-energy tensor of a minimally coupled scalar field
is equivalent to that of a perfect fluid, thereby defining an energy density
and pressure for the scalar field.
This equivalence leads to the following relation between $\varphi_0$
and $\rho_0+p_0$:
$(\partial_{N}\varphi_0)^2 =a^2{\cal H}^{-2}(\rho_0+p_0)$.}
so that energy conservation holds in the background:
\begin{equation}
\partial_N \rho_0=-3(\rho_0 + p_0), \label{cons_energy}
\end{equation}
it follows that\footnote{For the first equation differentiate~\eqref{cons_energy} and
with respect to
$N$ and use the definition of $c_s^2$ expressed in terms of $N$.
For the second equation differentiate
$(\partial_{N}\varphi_0)^2 =a^2{\cal H}^{-2}(\rho_0+p_0)$
using the result of the first equation.
One also requires $\partial_{N} (a{\cal H}^{-1})=(1+q)(a{\cal H}^{-1}),$
which follows from~\eqref{w,c_s} and~\eqref{eta_tau}. The second
equality in~\eqref{C_varphi} depends on~\eqref{q,w.relation}.  }
\begin{subequations}  \label{C_box_special}
\begin{align}
C_{\rho}&= -3(1+c_s^2), \label{C_rho} \\
C_{\varphi}&= (1+q) - \sfrac32(1+c_s^2) = \sfrac32(w - c_s^2).\label{C_varphi}
\end{align}
\end{subequations}
Second, the remaining source terms $\Box_{rem,\bullet}$ have the following form for the different choices of ${\Box}$ in~\eqref{box_var}:
\begin{subequations}\label{rem}
\begin{align}
\psi_{rem,\bullet} &= {\mathbb D}_2(B)- {\mathbb D}_2(B_{\bullet}), \label{rem_psi} \\
\begin{split}
{\cal H}B_{rem,\bullet} &= (\partial_{N}+2q)\left({\mathbb D}_0({\cal H}B_\bullet )-
{\mathbb D}_0({\cal H}B)\right) \\& \qquad
+ 2{\cal S}^i\left[(\phi_{\mathrm p}+\phi_{\bullet}){\bf D}_i({\cal H}B_{\bullet})  -
(\phi_{\mathrm p}+ \phi){\bf D}_i({\cal H}B)\right], \label{rem_B}
\end{split}\\
{\cal H}V_{rem,\bullet} &=
2{\cal S}^i\left[ \phi_{\mathrm v}{\bf D}_i ({\cal H}V_{\bullet}-{\cal H}V)\right], \label{rem_V}\\
{\rho}_{rem,\bullet} &= 0,  \label{rem_delta} \\
{\varphi}_{rem,\bullet} &= 0,
\end{align}
\end{subequations}
where the subscripts $_{\mathrm p}$ and $_{\mathrm v}$ stands for
the Poisson and total matter gauge, respectively, which
are defined in section~\ref{change_of_gauge}. The
scalar mode extraction operator ${\cal S}^i$ and the
spatial differential operators\footnote{We decided to introduce
this shorthand notation because these expressions occur frequently in
second order cosmological perturbation theory.
See appendix~\ref{spat.diff.op} for some notational motivation and
historical background concerning these expressions.}
${\mathbb D}_0$ and ${\mathbb D}_2$
that appear in equations~\eqref{rem}
are defined in equations~\eqref{modeextractop}
and~\eqref{D0,2} in appendix~\ref{spat.diff.op}.
We note that the temporal gauge on the right side of equations~\eqref{rem}
is unspecified and can be chosen to be one of the standard gauges.

At this stage we introduce the normalized density perturbation ${}^{(r)}\!{\bdelta}$ according to
\begin{equation}
{}^{(r)}\!{\bdelta} =
\frac{{}^{(r)}\!\rho}{(-\sfrac13\partial_N{\rho}_0)},\qquad r=1,2, \label{def_delta1}
\end{equation}
which means that the kernel $\Box$ that is associated
with the density perturbation in~\eqref{box_var} is given by
\begin{equation}
{}^{(r)}\!{\Box}=\sfrac13 {}^{(r)}\!{\bdelta}.  \label{box_var_rho}
\end{equation}
The factor of $\sfrac13$ is included so that if conservation
of energy holds in the background
($\partial_N \rho_0=-3(\rho_0 + p_0)$) then~\eqref{def_delta1} becomes
\begin{equation}
{}^{(r)}\!{\bdelta} = \frac{{}^{(r)}\!\rho}{\rho_0 + p_0}.
\end{equation}
In this case the usual fractional perturbed energy density
${}^{(r)}\!\delta$ is easily obtained via
\begin{equation}
{}^{(r)}\!\delta = \frac{{}^{(r)}\!\rho}{\rho_0} = (1+w){}^{(r)}\!{\bdelta}.
\end{equation}

For convenience we now explicitly list the normalized
source-compensated second order variables given by~\eqref{hat_box},
where the kernels $\Box$ are given by~\eqref{box_var} and~\eqref{box_var_rho}:
\begin{subequations} \label{hat_variables}
\begin{align}
{}^{(2)}\!{\hat{\psi}} &=  {}^{(2)}\!{\psi} + 2{}^{(1)}\!{\psi}^2,  \\
{\cal H}{}^{(2)}\!{\hat{B}} &=
{\cal H}{}^{(2)}\!{B} + (1+q)({\cal H} {}^{(1)}\!{B})^2, \\
{\cal H}{}^{(2)}\!{\hat{V}} &=
{\cal H}{}^{(2)}\!{V} + (1+q)({\cal H} {}^{(1)}\!{V})^2, \\
\sfrac13{}^{(2)}\!{\hat{\bdelta}} &=
\sfrac13{}^{(2)}\!{\bdelta} - 3(1+c_s^2) \left(\sfrac13{}^{(1)}\!{\bdelta}\right)^2,
\label{delta_hat} \\
\lambda{}^{(2)}\!{\hat{\varphi}} &=
\lambda{}^{(2)}\!{\varphi}+ \sfrac32(w - c_s^2)(\lambda{}^{(1)}\!{\varphi})^2.
\label{hat_varphi}
\end{align}
\end{subequations}
Here we have introduced the notation
\begin{equation}
\lambda= - (\partial_N{\varphi}_0)^{-1},  \label{def_lambda}
\end{equation}
for the scale factor associated with the scalar field in equation~\eqref{box_var}.
We note that equations~\eqref{delta_hat} and ~\eqref{hat_varphi}
depend on the conservation of energy in the background.

One feature of equation~\eqref{trans_gen2_repeat} requires
comment. In this equation one can replace
${}^{(1)}\!{\Box}_{\bullet}$ on the right side by ${}^{(1)}\!{\Box}$
using~\eqref{trans_gen1_repeat}, and then
modify the definition of ${}^{(2)}\!\hat{\xi}^N_{\bullet}$ by
changing the sign in~\eqref{hat_z}.
Although this form of the equation may look more natural,
it turns out that the form we have given is
more convenient when we actually apply the equation to make
a change of gauge in section~\ref{change_of_gauge}.

\emph{Equation~\eqref{trans_gen2_repeat} with~\eqref{hat_variables}
and~\eqref{rem}  forms the first main result of this paper}
and provides the basis for the change of gauge formulas
in section~\ref{change_of_gauge}.
For ease of reference we now write out the unified formula~\eqref{trans_gen2_repeat}
with $\Box$ having the values in~\eqref{box_var} and~\eqref{box_var_rho}:
\begin{subequations} \label{trans_var2}
\begin{align}
{}^{(2)}\!{\hat{\psi}}_{\bullet} &=  {}^{(2)}\!{\hat{\psi}} -
{}^{(2)}\!\hat{\xi}^N_{\bullet} +
2{}{}^{(1)}\!\xi^N_{\bullet}\, \partial_{N}\! {}^{(1)}\!{\psi}_{\bullet}  +
\psi_{rem, {\bullet}}. \label{trans_psi2}  \\
{\cal H}{}^{(2)}\!{\hat B}_{\bullet} &=  {\cal H}{}^{(2)}\!{\hat B} -
{}^{(2)}\!\hat{\xi}^N_{\bullet} +
2{}^{(1)}\!\xi^N_{\bullet}\, \partial_{N} ({\cal H}{}^{(1)}\!B_{\bullet})  +
{\cal H}B_{rem,\bullet}, \label{trans_B2}  \\
{\cal H}{}^{(2)}\!{\hat V}_{\bullet} &=  {\cal H}{}^{(2)}\!{\hat V} -
{}^{(2)}\!\hat{\xi}^N_{\bullet} +
2{}{}^{(1)}\!\xi^N_{\bullet}\, \partial_{N} ({\cal H}{}^{(1)}\!{V}_{\bullet}) +
{\cal H}V_{rem,\bullet}, \label{trans_V2}  \\
\sfrac13{}^{(2)}\!{\hat{\bdelta}}_{\bullet} &=  \sfrac13{}^{(2)}\!{\hat{\bdelta}}
 - {}^{(2)}\!\hat{\xi}^N_{\bullet} +
2{}{}^{(1)}\!\xi^N_{\bullet}\, \partial_{N}\! \left(\sfrac13{}^{(1)}\!{\bdelta}_{\bullet}\right),
\label{trans_delta2}  \\
\lambda{}^{(2)}\!{\hat{\varphi}}_{\bullet} &= \lambda {}^{(2)}\!{\hat{\varphi}} -
{}^{(2)}\!\hat{\xi}^N_{\bullet} +
2{}{}^{(1)}\!\xi^N_{\bullet}\, \partial_{N}(\lambda {}^{(1)}\!{\varphi}_{\bullet}),
\label{trans_varphi2}
\end{align}
\end{subequations}
where $\lambda$ is given by~\eqref{def_lambda}
and the remainder terms by~\eqref{rem}.
However, we need to augment this set of
equations with transformation equations for the metric  variable $\phi$,
which has to be treated separately since its
transformation law involves the time derivative of the gauge field.
At first order we have:
\begin{subequations} \label{trans_phi}
\begin{equation}
{}^{(1)}\!{\phi}_{\bullet} = {}^{(1)}\!{\phi} +
(\partial_{N} + 1 + q)({}^{(1)}\!{\xi}^N_{\bullet}), \label{trans_phi1}
\end{equation}
and at second order,
\begin{equation}\label{trans_phi2}
{}^{(2)}\!{\hat{\phi}}_{\bullet} = {}^{(2)}\!{\hat{\phi}} +
(\partial_{N} + 1 + q)({}^{(2)}\!{\hat{\xi}}^N_{\bullet}) +
2{}^{(1)}\!{\xi}^N_{\bullet}\, \partial_{N}\! {}^{(1)}\!{\phi}_{\bullet} + \phi_{rem,\bullet},
\end{equation}
where
\begin{align}
{}^{(2)}\!{\hat{\phi}} &=  {}^{(2)}\!{\phi} - 2{}^{(1)}\!{\phi}^2,\\
\phi_{rem,\bullet} &= \left(\partial_{N} {}^{(1)}\!{\xi}^N_{\bullet} \right)^2
- (\partial_{N} q)( {}^{(1)}\!{\xi}^N_{\bullet}) ^2.
\end{align}
\end{subequations}
We end this section by noting that there are other ways
defining the curvature perturbation $\psi$. In this paper we write the scalar
part of the perturbed spatial metric as $(1-2\psi)\delta_{ij}$ (which
we refer to as the Malik-Wands form, see for
example Malik and Wands (2009)~\cite{malwan09}),
while another choice is an exponential form ${\mathrm e}^{-2\psi_{SB}}\delta_{ij}$
first introduced by Salopek and Bond (1990)~\cite{salbon90} (see for example,
Lyth and Rodriguez (2005)~\cite{lytrod05}, section IIB). Equating the two forms,
Taylor expanding the exponential and performing a perturbation
expansion for $\psi_{SB}$ yields ${}^{(1)}\!\psi_{SB}={}^{(1)}\!\psi$
and ${}^{(2)}\!\psi_{SB}={}^{(2)}\!\psi +2({}^{(1)}\!\psi)^2$, showing
that $ {}^{(2)}\!\psi_{SB}={}^{(2)}\!{\hat \psi}$. We refer to section 2.1 in
Carrilho and Malik (2015)~\cite{carmal15} for two other possibilities.

\section{Performing  a change of gauge \label{change_of_gauge}}

Having fixed the spatial gauge (see Eq.~\eqref{Ccond}), we can now choose a temporal gauge to second order by
setting to zero the first and second perturbations of one the variables $B$, $\psi$, $V$, ${\bdelta}$, ${\varphi}$,
thereby specifying the gauge uniquely. We use the following terminology and subscripts to label the gauges:
\begin{itemize}
\item[i)] Poisson gauge, subscript ${}_{\mathrm p}$,
defined by $B_{\mathrm p} = 0$,
\item[ii)] uniform curvature gauge, subscript ${}_{\mathrm c}$,
defined by $\psi_{\mathrm c} = 0$,
\item[iii)] total matter gauge, subscript ${}_{\mathrm v}$,
defined by $V_{\mathrm v} = 0$,
\item[iv)] uniform density gauge,  subscript ${}_{\rho}$,
defined by ${\bdelta}_{\rho} = 0$,
\item[v)] uniform scalar field gauge,\footnote
{This gauge is naturally only available in a perturbed universe
with a scalar field. In this context it is in fact equivalent to the
total matter gauge, but it is helpful to give it a separate name.
This equivalence is established in a subsequent paper~\cite{uggwai18}.}
subscript ${}_{\mathrm {sc}}$,
defined by $\varphi_{\mathrm {sc}} = 0$.
\end{itemize}
In order to introduce a specific gauge labelled by ${}_\bullet$ we must determine the transition function
${}^{(2)}\!{\hat{\xi}}^N_{\bullet}$ using equations~\eqref{trans_var2}.
(We will not list the expressions for ${}^{(1)}\!\xi^N_{\bullet}$ below since they can easily be read off
from the second order equations: replace ${}^{(2)}$ by ${}^{(1)}$, omit the hats and drop the ${}_{rem}$ terms.)
Referring to the above definition of the gauges we choose
${}_{\bullet}={}_{\mathrm p}$ in~\eqref{trans_B2},
${}_{\bullet}={}_{\mathrm v}$ in~\eqref{trans_V2},
${}_{\bullet}={}_{\mathrm c}$ in~\eqref{trans_psi2},
${}_{\bullet}={}_{\rho}$ in~\eqref{trans_delta2} and
${}_{\bullet}={}_{\mathrm sc}$ in~\eqref{trans_varphi2},
to obtain the following results:
\begin{subequations} \label{solve_xi2}
\begin{align}
{}^{(r)}\!B_{\mathrm p} &= 0 \quad \implies \quad
{}^{(2)}\!{\hat{\xi}}^N_{\mathrm p} = {\cal H}{}^{(2)}\!{\hat B} +
{\cal H}B_{rem, {\mathrm p}},   \\
{}^{(r)}\!V_{\mathrm v} &= 0 \quad \implies \quad
{}^{(2)}\!{\hat{\xi}}^N_{\mathrm v} = {\cal H}{}^{(2)}\!{\hat V} +
{\cal H}V_{rem, {\mathrm v}},   \\
{}^{(r)}\!\psi_{\mathrm c} &= 0 \quad \implies \quad
{}^{(2)}\!{\hat{\xi}}^N_{\mathrm c} = {}^{(2)}\!{\hat \psi} +
\psi_{rem, {\mathrm c}},   \\
{}^{(r)}\!{\bdelta}_{\rho} &=0 \quad \implies \quad
{}^{(2)}\!{\hat{\xi}}^N_{\rho} = \sfrac13{}^{(2)}\!{\hat \bdelta}, \\
{}^{(r)}\!\varphi_{\mathrm {sc}} &= 0 \quad \implies \quad
{}^{(2)}\!{\hat{\xi}}^N_{\mathrm {sc}} = \lambda{}^{(2)}\!{\hat \varphi}.
\end{align}
\end{subequations}
\emph{These expressions for the gauge fields at second order
represent the second main result of this paper.} Their
concise form is a consequence of using the hatted variables.

The final step is to successively substitute the expressions~\eqref{solve_xi2}
into~\eqref{trans_gen2_repeat}. This immediately gives the following
change of gauge formulas at second order:
\begin{subequations} \label{box_change2}
\begin{align}
{}^{(2)}\!\hat{\Box}_{\mathrm p} &= {}^{(2)}\!\hat{\Box} - {\cal H}{}^{(2)}\!\hat {B} +
2{\cal H}{}^{(1)}\!B\,\partial_N {}^{(1)}\!\Box_{\mathrm p} +
\Box_{rem,{\mathrm p}} - {\cal H}B_{rem, {\mathrm p}},  \label{box_change_p2} \\
{}^{(2)}\!\hat{\Box}_\mathrm{v} &= {}^{(2)}\!\hat{\Box} - {\cal H}{}^{(2)}\!\hat {V} +
2{\cal H}{}^{(1)}\!V\,\partial_N {}^{(1)}\!\Box_\mathrm{v}  +
\Box_{rem,\mathrm{v}} -{\cal H}V_{rem, {\mathrm v}},  \label{box_change_v2} \\
{}^{(2)}\!{\hat{\Box}}_{\mathrm c} &=  {}^{(2)}\!{\hat{\Box}} - {}^{(2)}\!\hat{\psi} +
2{}^{(1)}\!\psi\,\partial_N {}^{(1)}\!{\Box}_{\mathrm c} +
\Box_{rem,{\mathrm c}} - \psi_{rem, {\mathrm c}}, \label{box_change_c2} \\
{}^{(2)}\!{\hat{\Box}}_{\rho} &=  {}^{(2)}\!{\hat{\Box}} - \sfrac13{}^{(2)}\!\hat{\bdelta} +
2(\sfrac13{}^{(1)}\!\bdelta)\,\partial_N {}^{(1)}\!{\Box}_{\rho} +
\Box_{rem,{\rho}}, \label{box_change_rho2} \\
{}^{(2)}\!{\hat{\Box}}_{\mathrm sc}&=  {}^{(2)}\!{\hat{\Box}} - \lambda{}^{(2)}\!\hat{\varphi} +
2(\lambda{}^{(1)}\!\varphi)\,\partial_N {}^{(1)}\!{\Box}_{\mathrm sc} +
\Box_{rem,{\mathrm sc}},  \label{box_change_sc2}
\end{align}
\end{subequations}
where the $\Box_{rem,\bullet}$ terms are given by~\eqref{rem},
and $\Box$ represents any of the symbols in equation~\eqref{box_var}
and~\eqref{box_var_rho}. The gauge on the right side is unspecified and can be
chosen to be one of the standard gauges. For example if one wishes
to transform from the total matter gauge to the
uniform curvature gauge, one would use the third
equation with subscripts ${}_{\mathrm v}$ added on the right side:
\begin{equation}
{}^{(2)}\!\hat{\Box}_{\mathrm c} = {}^{(2)}\!\hat{\Box}_{\mathrm v} -
{}^{(2)}\!\hat {\psi}_{\mathrm v} +
2{}^{(1)}\!\psi_{\mathrm v}\,\partial_N {}^{(1)}\!\Box_{\mathrm c} +
\Box_{rem,{\mathrm c},{\mathrm v}} - \psi_{rem, {\mathrm c},{\mathrm v}},
\end{equation}
with
\begin{equation}
{}^{(1)}\!\Box_{\mathrm c} = {}^{(1)}\!\Box_{\mathrm v} -
{}^{(1)}\!\psi_{\mathrm v} ,
\end{equation}
at first order.\footnote{As with equations~\eqref{solve_xi2}, the first order
formulas can be read off from the second order formulas by inspection, since they correspond
to the leading order terms.}
The remainder terms are obtained from equations~\eqref{rem} by choosing
the total matter gauge on the right side. In the present example we obtain
\begin{equation}
\psi_{rem, {\mathrm c},{\mathrm v}}=
{\mathbb D}_2(B_{\mathrm v})- {\mathbb D}_2(B_{\mathrm c}).
\end{equation}
In summary \emph{equation~\eqref{box_change2}, in conjunction with
the definition~\eqref{hat_variables} of the hatted variables,
represent the main goal of this paper.}
They provide an efficient algorithm for calculating any
of the gauge invariants in any of the five gauges, as illustrated
in the next section. Although our primary motivation was to
simplify and unify the change of gauge formulas at second order,
we note that equations~\eqref{box_change2} also
give a useful overview of the situation at linear order. By replacing
${}^{(2)}$ with ${}^{(1)}$ and by omitting the hats and dropping the ${}_{rem}$ terms
one can read off familiar relations such as
\begin{equation}
\psi_{\mathrm v}=\psi_{\mathrm p}-{\cal H}V_{\mathrm p},
\quad \psi_{\mathrm p}=-{\cal H}B_{\mathrm c}, \quad
\bdelta_{\mathrm v} =\bdelta_{\mathrm p} - 3{\cal H}V_{\mathrm p}, \quad
\psi_{\rho}=-\sfrac13\bdelta_{\mathrm c},
\quad \psi_{sc}=-\lambda \varphi_{\mathrm c}.
\end{equation}
Finally, we recall that the change of gauge formulas for the metric perturbation $\phi$
have to be treated separately and are given by equations~\eqref{trans_phi},
with the specific gauge field to be obtained from equations~\eqref{solve_xi2}
once the two gauges have been chosen. An example at linear order is
\begin{equation}
{}^{(1)}\!{\phi}_{\mathrm v} = {}^{(1)}\!{\phi}_{\mathrm c} +
(\partial_{N} + 1 + q)({}^{(1)}\!{\xi}^N_{\mathrm v,\mathrm c}), \quad
\text {with}\quad {}^{(1)}\!{\xi}^N_{\mathrm v,\mathrm c}={\cal H}V_{\mathrm c}.
\end{equation}
%

\section{Examples \label{applications} }

In this section we give examples of using the general
equations~\eqref{box_change2} to calculate
second order gauge invariants of interest in current
research in cosmology. The expressions we obtain are
more concise than those in the literature because of
our use of the hatted variables and the
differential operators ${\mathbb D}_0$ and ${\mathbb D}_2$.
The latter feature, in particular, simplifies the
representation of the terms involving spatial derivatives. In order to make
comparisons with the literature it is necessary to
expand our expressions by using
the definition~\eqref{hat_variables} of the hatted
variables and the definition~\eqref{D0,2} of
${\mathbb D}_0$ and ${\mathbb D}_2$. The latter
definitions lead to the following identities
that will be useful when making comparisons:
\begin{subequations} \label{D0,2_diff}
\begin{align}
{\mathbb D}_0(A) -{\mathbb D}_0(B) &=
{\cal S}^{ij}[{\bf D}_i(A+B){\bf D}_j(A-B)], \label{D0_diff} \\
{\mathbb D}_2(A) -{\mathbb D}_2(B) &= \sfrac13(
{\bf D}^2{\cal S}^{ij} - \delta^{ij})[{\bf D}_i(A+B){\bf D}_j(A-B)],  \label{D2_diff}
\end{align}
\end{subequations}
where the scalar mode extraction operator
${\cal S}^{ij}$ is defined by~\eqref{modeextractop}.
We will frequently change from $e$-fold time $N$ to conformal time
$\eta$ using equation~\eqref{eta_tau} in order to make comparisons
with the literature.
We note the process of expanding our expressions to make comparisons
with the literature as illustrated in section~\ref{applications} and in
appendix~\ref{comp_lit}, can be tedious. We regard this as a measure of how
concise our expressions are, and we emphasize that this is not
something one has to do when using our formalism in practice,
as discussed in section~\ref{discussion}.
A bonus of the conciseness is that
it is easier to avoid errors and to find simpler unifying expressions.

Our first example concerns ${}^{(2)}\!\psi_{\rho}$, the second order curvature
perturbation in the uniform density gauge, which is
important as a conserved quantity on
super-horizon scales.\footnote{See, for example,
Bartolo \emph{et al} (2010)~\cite{baretal10}, equation (36).}
Choosing $\Box=\psi$ in~\eqref{box_change_rho2} and using~\eqref{rem_psi}
for the remainder term immediately gives
\begin{equation}
{}^{(2)}\!\hat{\psi}_{\rho} = {}^{(2)}\!\hat{\psi} -\sfrac13 {}^{(2)}\!\hat{\bdelta} +
\sfrac23{}^{(1)}\!{\bdelta}\partial_N {}^{(1)}\!\psi_{\rho}  -
{\mathbb D}_2({}^{(1)}\!B_{\rho}) + {\mathbb D}_2({}^{(1)}\!B).  \label{psi_rho.2}
\end{equation}
Equation~\eqref{psi_rho.2} is a concise version of equation (7.71) in Malik and Wands
(2009)~\cite{malwan09}.\footnote{Set $E_1=0$ in their equation to fix the spatial gauge.
There are a number of typos.} If we choose the arbitrary temporal gauge to be the
uniform curvature gauge ($\psi_{\mathrm c}=0)$ equation~\eqref{psi_rho.2} becomes
\begin{equation}
{}^{(2)}\!\hat{\psi}_{\rho} = -\sfrac13 {}^{(2)}\!\hat{\bdelta}_{\mathrm c} +
\sfrac23{}^{(1)}\!{\bdelta}_{\mathrm c}\partial_N {}^{(1)}\!\psi_{\rho}  -
{\mathbb D}_2({}^{(1)}\!B_{\rho}) + {\mathbb D}_2({}^{(1)}\!B_{\mathrm c}),
\end{equation}
which is a concise version of equation (3.3) in Christofferson \emph{et al} (2015)~\cite{chretal15},
which relates ${}^{(2)}\!{\psi}_{\rho}$ to ${}^{(2)}\!{\bdelta}_{\mathrm c}$,
the second order density perturbation in the uniform curvature gauge. To compare with
earlier literature we change from $N$ to $\eta$ and make the replacement
${}^{(r)}\!\bdelta = -3({\cal H}/\rho_0'){}^{(r)}\!\rho$, which leads to
\begin{subequations}  \label{psi_rho}
\begin{equation}
{}^{(2)}\!\psi_{\rho}= \frac{\cal H}{\rho_{0}'} {}^{(2)}\!\rho_{\mathrm c} -
\frac{\cal H}{(\rho_0' )^2}{}^{(1)}\!\rho_{\mathrm c}\left(2{}^{(1)}\!\rho_{\mathrm c}'+
(5+3c_s^2){\cal H}{}^{(1)}\!\rho_{\mathrm c}\right) -
{\mathbb D}_2({}^{(1)}\!B_{\rho}) + {\mathbb D}_2({}^{(1)}\!B_{\mathrm c}),
\end{equation}
where
\begin{equation}
{\mathbb D}_2({}^{(1)}\!B_{\rho}) - {\mathbb D}_2({}^{(1)}\!B_{\mathrm c})
= \sfrac{1}{27}{\cal H}^{-2}({\bf D}^2{\cal S}^{ij} - \delta^{ij})
[{\bf D}_i({}^{(1)}\!\bdelta_{\mathrm c} -
6{\cal H}{}^{(1)}\!B_{\mathrm c}){\bf D}_j{}^{(1)}\!\bdelta_{\mathrm c}],
\end{equation}
\end{subequations}
the latter relation following from~\eqref{D2_diff} and
${\cal H}{}^{(1)}\!B_{\rho}= {\cal H}{}^{(1)}\!B_{\mathrm c} -
\sfrac13{}^{(1)}\!\bdelta_{\mathrm c}$.
We find that equation (3.3) in~\cite{chretal15}
agrees with~\eqref{psi_rho}.\footnote{In rearranging
the $O({\bf D}^2)$ terms one has to use~\eqref{def_delta1},
and the definition of ${\cal S}_{ij}$.}
Equation~\eqref{psi_rho} has also been given by Carrilho and
Malik (2015)~\cite{carmal15} (see equation (3.3)).

Our second example concerns the density perturbation at second
order in the total matter gauge
${\bdelta}_{\mathrm v}$ which is used when deriving the generalized
Poisson equation in second order
perturbation theory in relativistic cosmology (see Hidalgo
\emph{et al} (2013)~\cite{hidetal13}).
Choosing $\Box= \sfrac13\bdelta$ in~\eqref{box_change_v2} and
using~\eqref{rem_delta} for the remainder term
we express ${\bdelta}_{\mathrm v}$ in terms of an arbitrary temporal gauge:
\begin{equation}
{}^{(2)}\!\hat{\bdelta}_{\mathrm v} = {}^{(2)}\!\hat{\bdelta} -
3{\cal H}{}^{(2)}\!\hat{V} + 2{\cal H}{}^{(1)}\!V\partial_{N} {}^{(1)}\!{\bdelta}_{\mathrm v} +
6{\cal S}^i[{}^{(1)}\!\phi_{\mathrm v}{\bf D}_i {\cal H}{}^{(1)}\!V],  \label{delta_change.2}
\end{equation}
where the mode extraction operator ${\cal S}^i$ is defined in equation~\eqref{modeextractop}.

Expanding our equation, changing from $N$ to $\eta$ and using~\eqref{def_delta1} leads to
\begin{equation}
{}^{(2)}\!\rho_{\mathrm v} = {}^{(2)}\!\rho +3\rho_0'{}^{(2)}\!{V} +
 {}^{(1)}\!V[2{}^{(1)}\!\rho_{\mathrm v}' +
3\rho_0'(1+c_s^2 +\sfrac12(1+w)){\cal H}{}^{(1)}\!V] -
2\rho_0'{\cal S}^i[{}^{(1)}\!\phi_{\mathrm v}{\bf D}_i {}^{(1)}\!V].
\label{delta_change.3}
\end{equation}
Equation (3.10) in~\cite{hidetal13} can be simplified to have the form~\eqref{delta_change.3}
(subject to a few differing coefficients) when we choose the arbitrary temporal gauge
on the right side of~\eqref{delta_change.3} to be the Poisson gauge. To make the comparison it is easier to
perform further manipulations on~\eqref{delta_change.3} and transform it into the Hidalgo \emph{et al}
form.\footnote{Use ${}^{(1)}\!\phi_{\mathrm v} = {}^{(1)}\!\phi +{}^{(1)}\!V' +{\cal H}{}^{(1)}\!V$,
introduce ${}^{(r)}\!\delta = {}^{(r)}\!\rho/\rho_0$, use $2{\cal S}^i[V{\bf D}_i V] = V^2$,
replace $1+c_s^2$ using $w' = 3{\cal H}(1+w)((1+w)-(1+c_s^2))$, and assume conservation of energy
$\rho_0' = -3{\cal H}(1+w)\rho_0$.}

As our third example, by choosing $\Box = \sfrac13\bdelta$
in~\eqref{box_change_c2} and
using~\eqref{rem_delta} for the remainder term, we
express ${\bdelta}_{\mathrm c}$ in terms of an
arbitrary temporal gauge:
\begin{equation}\label{delta_change_c2}
{}^{(2)}\!{\hat{\bdelta}}_{\mathrm c}=  {}^{(2)}\!{\hat{\bdelta}} - 3{}^{(2)}\!\hat{\psi} +
2{}^{(1)}\!\psi\partial_N {}^{(1)}\!\bdelta_{\mathrm c} +
3({\mathbb D}_2({}^{(1)}\!B_{\mathrm c}) - {\mathbb D}_2({}^{(1)}\!B)).
\end{equation}
Expanding our equation, changing from $N$ to $\eta$ and using~\eqref{def_delta1} leads to
\begin{subequations}\label{delta_c2_expand}
\begin{equation}
\begin{split}
{}^{(2)}\!{\rho}_{\mathrm c} &=  {}^{(2)}\!\rho +
\frac{\rho_0' }{\cal H}{}^{(2)}\!\psi  + \frac{{}^{(1)}\!\psi}{\cal H}
\left[2{}^{(1)}\!\rho' + \frac{\rho_0' }{\cal H}\left(2{}^{(1)}\!\psi'  -
{\cal H}(1+3c_s^2){}^{(1)}\!\psi \right)\right] \\
& \quad -
\frac{\rho_0' }{\cal H}({\mathbb D}_2({}^{(1)}\!B_{\mathrm c}) - {\mathbb D}_2({}^{(1)}\!B)),
\end{split}
\end{equation}
where
\begin{equation}
{\mathbb D}_2({}^{(1)}\!B_{\mathrm c}) - {\mathbb D}_2({}^{(1)}\!B)=
\sfrac13 {\cal H}^{-2} ({\bf D}^2{\cal S}^{ij} - \delta^{ij})
[{\bf D}_i({}^{(1)}\!\psi - 2{\cal H}{}^{(1)}\!B){\bf D}_j {}^{(1)}\!\psi],
\end{equation}
\end{subequations}
the latter relation following from~\eqref{D2_diff} and
${\cal H}{}^{(1)}\!B_{\mathrm c}={\cal H}{}^{(1)}\!B-{}^{(1)}\!\psi$.
Equation~\eqref{delta_c2_expand} agrees with equation (7.35) in Malik and Wands
(2009)~\cite {malwan09} with the gauge fixed so that $E_1=0$.\footnote{To make the comparison
note that ${\cal H}\rho_0''- \rho_0' {\cal H}'= -3 \rho_0' {\cal H}^2(1+c_s^2)$,
and write the spatial derivative terms using our notation ${\bf D}_i, {\bf D}^2$ and ${\cal S}^{ij}$.}

Our final example in this section concerns the curvature perturbation
in the uniform scalar field gauge
${}^{(2)}\!\hat{\psi}_{\mathrm sc}$ which is a conserved quantity on super-horizon scales
in a scalar field dominated universe (Vernizzi (2005)~\cite{ver05}). Choosing $\Box=\psi$
in~\eqref{box_change_sc2} and using~\eqref{rem_psi} for the remainder term immediately gives
\begin{equation}
{}^{(2)}\!\hat{\psi}_{\mathrm sc} = {}^{(2)}\!\hat{\psi} - \lambda{}^{(2)}\!\hat{\varphi} +
2\lambda{}^{(1)}\!{\varphi}\partial_N {}^{(1)}\!\psi_{\mathrm sc}  -
{\mathbb D}_2({}^{(1)}\!B_{\mathrm sc}) + {\mathbb D}_2({}^{(1)}\!B).  \label{psi_sc.2}
\end{equation}
Expanding our equation leads to\footnote{Here we have used
equation~\eqref{hat_box} for $\lambda{}^{(2)}\!\hat{\varphi}$
and equation~\eqref{C_box} for $C_{\varphi}$, as well as the
first order relation ${}^{(1)}\!\psi_{\mathrm sc} =
{}^{(1)}\!\psi - \lambda{}^{(1)}\!{\varphi}.$}
\begin{equation}
\begin{split}
{}^{(2)}\!{\psi}_{\mathrm sc} = {}^{(2)}\!\psi - \lambda{}^{(2)}\!\varphi +&
\lambda {}^{(1)}\!{\varphi}\!\left[ - 2\lambda \partial_{N}\!{}^{(1)}\!{\varphi} -
(\lambda \partial_{N}^2{\varphi}_0 + 2)\lambda{}^{(1)}\!{\varphi} +
2(\partial_{N} +2) {}^{(1)}\!\psi)\right]   \\& -
{\mathbb D}_2({}^{(1)}\!B_{\mathrm sc}) + {\mathbb D}_2({}^{(1)}\!B),
\end{split}
\end{equation}
with $\lambda =  - ({\partial_N{\varphi}}_0)^{-1}.$ Converting to $\eta$ as time variable yields
equation (30) in~\cite{ver05}, when specialized to the long wavelength limit.\footnote{Note
that ${\cal H}^2(\lambda \partial_N^2{\varphi}_0 + 2) =
\lambda  {\varphi}_0'' + {\cal H}' + 2{\cal H}^2$, where
$\lambda = - (\partial_N{\varphi_0})^{-1} = - {\cal H}({\varphi_0'})^{-1}$ in our notation.
However, Vernizzi uses the convention $\dot{f} = \partial_{\eta}$. In addition, since Vernizzi restricts
consideration to long wavelength perturbations the terms
${\mathbb D}_2(B_{\mathrm sc}) - {\mathbb D}_2(B)$, which are $O({\bf D}^2)$, do not appear.}
We note in passing that this transformation formula plays a central role in finding a
conserved quantity and explicit solutions at second order, as discussed in the next
section and in detail in the follow up papers~\cite{uggwai19b} and~\cite{uggwai19c},
called UW3 and UW4, respectively, below.

\section{Discussion~\label{discussion}}

The present paper is the first of four closely connected
papers dealing with scalar perturbations up to second order.
In the present paper, which we will refer to as UW1, we have introduced
new second order variables and a new second order gauge vector field,
which simplifies the change of gauge formulas at
second order, and provides an efficient unifying algorithm
for calculating any gauge invariant in the commonly used gauges.
This is important since it is often desirable to use several
gauges when addressing a given problem. In the second paper, called
UW2~\cite{uggwai18}, we present five ready-to-use systems of
governing equations for second order perturbations. These two papers
constitute the foundation for subsequent physical applications,
illustrated by UW3~\cite{uggwai19b} and UW4~\cite{uggwai19c}.

In UW3 we use the new variables and gauge transformation formulas,
and apply them to the equations given in UW2 to produce new dimensionless
gauge-invariant conserved quantities and explicit general solutions,
containing both the so-called growing and decaying modes, for second order
perturbations in the super-horizon regime. This is made possible
due to that the dimensionless source-compensated ``hatted''
second order perturbation variables simplify the analysis of perturbations in the
super-horizon regime in a number of ways. For example,
in this regime the perturbed energy conservation equations
can be written in the following form in terms of hatted variables:
\begin{subequations}  \label{cons_energy_SH}
\begin{align}
\partial_{N}({}^{(1)}\!{\bdelta} - 3{}^{(1)}\!{\psi}) &\approx
-3{}^{(1)}\!{\Gamma}, \label{cons_energy1_SH}  \\
\partial_{N}({}^{(2)}\!\hat{\bdelta} - 3{}^{(2)}\!\hat{\psi}) &\approx
-3({}^{(2)}\!{\Gamma} - 2{}^{(1)}\!\Gamma^2 )+
2\partial_{N}({}^{(1)}\!\Gamma{}^{(1)}\!{\bdelta}),
\label{cons_energy2_SH}
\end{align}
\end{subequations}
with a simple quadratic source term in the second
order equation (see UW3~\cite{uggwai19b}).
Here ${}^{(r)}\!\Gamma,\, r=1,2$, are the
non-adiabatic pressure perturbations (see UW2~\cite{uggwai18}).
It follows from~\eqref{cons_energy_SH}, specialized to the
uniform curvature gauge (${}^{(r)}\!{\psi}=0,\,r=1,2$)
that ${}^{(1)}\!{\bdelta}_{\mathrm c}$ and the \emph{hatted} second order perturbation
${}^{(2)}\!\hat{\bdelta}_{\mathrm c}$ are conserved for
adiabatic perturbations (${}^{(r)}\!\Gamma=0$, $r=1,2$)
in the super-horizon regime. Note, however, that the
\emph{unhatted} second order density perturbation
${}^{(2)}\!{\bdelta}_{\mathrm c}$ is \emph{not} conserved unless $c_s^2$
is constant, as follows from~\eqref{delta_hat}.
Another example is that when using the uniform curvature gauge
in the super-horizon regime, the perturbed Einstein equations assume a
particularly simple form when expressed in terms of hatted variables,
which leads to further conserved quantities.
In particular if the source is a minimally coupled scalar field with an
\emph{arbitrary} scalar field potential we obtain a new second
order conserved quantity for the scalar field, which is used in
UW4~\cite{uggwai19c} to obtain new physical results for scalar fields for second order
perturbations in the long wavelength limit, without imposing the slow-roll
approximation.

\begin{appendix}

\section{Relation with Malik and Wands (2009) \label{comp_lit}}

We begin by listing the main differences between
Malik and Wands (2009)~\cite{malwan09} and the present paper.
First, they give a more general framework for the gauge transformation formulas
than we do in that they do not require the perturbations at first order to be
purely scalar. Second, they do not use the gauge freedom to set $C=0$, $C_i=0$ as we have
done ($C$ and $C_i$ are labeled $E$ and $F_i$ in their paper). Their generating vector field
$\xi$ at first order, given by their equation (6.17),
\begin{equation}
\xi^{\mu}_1 = (\alpha_1, \partial^i\beta_1 +\gamma_1^i),
\end{equation}
is thus more general than ours, which is given by equation~\eqref{xi_restricted}
(also note that they use subscripts to denote the order of the perturbation).
As a result of these differences when comparing equations it is necessary to set
$E=0$, $F=0$, $S_i=0$, $h_{ij}=0$ at first order in the metric perturbations (their
equations (2.8)-(2.12)), and to set $\beta_1=0$, $\gamma_1^i=0$ in their generating
gauge vector field. Then identify $\alpha_1\equiv {}^{(1)}\xi^\eta=
{\cal H}^{-1}{}^{(1)}\xi^N$. Their generating
gauge vector field at second order is general,
$\xi^{\mu}_2 = (\alpha_2, \partial^i\beta_2 +\gamma_2^i)$. In accordance
with~\eqref{xi_restricted}, we identify
\begin{equation}
\alpha_2= {}^{(2)}\!\xi^\eta=
{\cal H}^{-1}{}^{(2)}\!\xi^N, \qquad
\beta_2\equiv {}^{(2)}\!\xi,\qquad
\gamma_2^i= {}^{(2)}\!{\tilde {\xi_i}}. \label{identify_gvf2}
\end{equation}

The third difference is in the treatment of the density perturbation. Malik and Wands
begin with the standard unscaled perturbation expansion (their equation (6.16)):
\begin{equation}
\rho =\rho_0 + \rho_1 + \sfrac12 \rho_2,
\end{equation}
and do not assume local energy conservation. We, on the other hand,
in several equations assume local energy conservation and then define the scaled perturbations
\begin{equation}
{}^{(r)}\!\bdelta = \frac{\rho_r}{\rho_0+p_0},  \qquad r=1,2.
\end{equation}
 However, normalizing with $\rho_0 + p_0$ is
equivalent to normalizing with $-\rho_0'/(3{\cal H})$, which suggests a natural
generalization in the case local energy conservation does not hold.

The fourth difference is in the treatment of the velocity perturbation.
Malik and Wands begin with the perturbation expansion (their equation (4.4)):
\begin{equation}
u^i = a^{-1}(v^i_1 + \sfrac12 v^i_2), \qquad
v^i_r =\partial^i v_r + {\tilde v}^j_r, \qquad r=1,2,
\end{equation}
of the contravariant spatial components of the 4-velocity, whereas we expand the covariant
components as in~\eqref{vel_exp}, using $V$ instead of $v$ as the scalar perturbation of the
velocity. It follows that $v$ and $V$ are related as follows:\footnote{Expand the relation
$u_a = g_{ab} u^b$. Here we are using the Malik and Wands subscript convention to indicate
the order of the perturbation.}
\begin{subequations} \label{v_V}
\begin{align}
V_1 &= v_1 + B_1,  \\
V_2 &= v_2 + B_2 - 2{\cal S}^i[\phi_1{\bf D}_iB_1 + 2\psi_1{\bf D}_i v_1].
\end{align}
\end{subequations}

We are using the same letters for the scalar metric perturbations, namely $(\phi,B,\psi)$ as
Malik and Wands. Their transformation laws for these variables are given by equations
(6.37)-(6.39) at first order and (6.47), (6.51) and (6.58) at second order. In order to
facilitate a comparison we write our formulas using the Malik and Wands variables and notation.
For ${}^{(2)}\!\phi$, given by equation~\eqref{trans_phi2}, we
use~\eqref{identify_gvf2} to obtain
\begin{equation}
{\tilde \phi_2} = \phi_2 + \alpha_2' +{\cal H}\alpha_2 + \alpha_1(\alpha_1'' +
5{\cal H}\alpha_1' + ({\cal H}'+2 {\cal H}^2)\alpha_1 + 2 \phi_1' + 4{\cal H}\phi_1 ) +
2\alpha_1' (\alpha_1' + 2\phi_1). \label{phi_MW}
\end{equation}
We have consistency with Malik and Wands equation (6.47). For ${}^{(2)}\!\psi$,
given by equation~\eqref{trans_psi2}, we
use~\eqref{identify_gvf2} to obtain
\begin{equation}
\begin{split}
{\tilde \psi_2} &= \psi_2 - {\cal H}\alpha_2 - \alpha_1\left({\cal H}\alpha_1' +
({\cal H}' + 2 {\cal H}^2)\alpha_1 - 2\psi_1'- 4{\cal H}\psi_1\right) \\
& + \sfrac13({\bf D}^2{\cal S}^{ij} - \delta^{ij})[{\bf D}_i(2B_1 - \alpha_1) {\bf D}_j  \alpha_1].
\end{split} \label{psi_MW}
\end{equation}
To obtain agreement with Malik and Wands we write their (6.58) in the form
\begin{equation}
{\tilde \psi_2}=\psi_2-{\cal H}\alpha_2 +
\sfrac16({\bf D}^2{\cal S}^{ij} - \delta^{ij})\chi_{ij}, \label{malwan1}
\end{equation}
where $\chi_{ij}$ is given by (6.54) specialized to scalar perturbations at first order by setting
$C_{1ij} = -\psi \delta_{ij}$, $B_{1i} = {\bf D}_iB$ and $\xi^k_1=0$, which yields
\begin{equation}
\begin{split}
\chi_{ij} = 2\alpha_1&\left({\cal H}\alpha_1' +
({\cal H}'+2 {\cal H}^2)\alpha_1 -2\psi_1'-4{\cal H}\psi_1\right) \delta_{ij}\\
&\qquad + 2({\bf D}_i B_1 {\bf D}_j\alpha_1 +{\bf D}_i  \alpha_1{\bf D}_jB_1 -
{\bf D}_i\alpha_1 {\bf D}_j  \alpha_1).
\end{split}   \label{malwan2}
\end{equation}
Substituting~\eqref{malwan2} in~\eqref{malwan1} yields~\eqref{psi_MW}.

For ${}^{(2)}\!B$, given by equation~\eqref{trans_B2}, we
use~\eqref{identify_gvf2} to obtain
\begin{equation}
\begin{split}
{\tilde B_2} &= B_2 - \alpha_2 + \beta_2' - \alpha_1(\alpha_1' + {\cal H}\alpha_1) \\
&- 2{\cal S}^i[-2\phi\,{\bf D}_i \alpha_1 + \alpha_1{\bf D}_i(B_1'+{\cal H}B_1) +
(\alpha_1' + {\cal H}\alpha_1){\bf D}_i(B_1 - \alpha_1)],
\end{split} \label{B_MW}
\end{equation}
where we have introduced the Malik-Wands notation $\beta_2={}^{(2)}\!\xi$, which is given by
\begin{equation}
\beta_2 = -{\cal S}^{ij}[{\bf D}_i(2B_1 - \alpha_1) {\bf D}_j  \alpha_1].  \label{beta_2}
\end{equation}
To obtain agreement with Malik and Wands we write their (6.50) in the form
\be {\tilde B_2}= B_2-\alpha_2 +\beta_2' + {\cal S}^i \chi_{Bi}, \label{malwan3} \ee
where $\beta_2$ is given by (6.59) with the spatial gauge fixed
so that ${\tilde E_2}=0=E_2$,
\begin{equation}
\beta_2 = -\sfrac12{\cal S}^{ij}\chi_{ij}, \label{malwan4}
\end{equation}
and where $\chi_{Bi}$ is given by (6.49) specialized to scalar perturbations at first order by setting
$B_{1i}={\bf D}_iB$, and $\xi^k_1=0$, which yields
\begin{equation}
\chi_{Bi} = - {\bf D}_i[\alpha_1 (\alpha_1'+{\cal H}\alpha_1) ]
 + 2[-2\phi\,{\bf D}_i \alpha_1 +\alpha_1{\bf D}_i (B_1'+{\cal H}B_1)+
(\alpha_1'+{\cal H}\alpha_1){\bf D}_i(B_1-\alpha_1)]. \label{malwan5}
\end{equation}
Note that~\eqref{malwan4} with~\eqref{malwan2} yields~\eqref{beta_2}.
Substituting~\eqref{malwan5} in~\eqref{malwan3} yields~\eqref{B_MW}.

We now consider the matter perturbations. For ${}^{(2)}\!V$ the ``Malik and Wands form''
of our equation~\eqref{trans_V2} is as follows:
\begin{equation}
{\tilde V_2} = V_2-\alpha_2 - \alpha_1(\alpha_1' + {\cal H}\alpha_1)
+ 2{\cal S}^i[-\phi \,{\bf D}_i \alpha_1 + \alpha_1{\bf D}_i(V_1' + {\cal H}V_1) ].
\label{V_MW}
\end{equation}
To compare with Malik and Wands we need to use equations~\eqref{v_V} and~\eqref{B_MW}
to obtain the transformation law for $v_2$:
\begin{equation}
{\tilde v}_2 = v_2 - \beta_2' + 2{\cal S}^i[\alpha_1 {\bf D}_i(v_1' - {\cal H}v_1)],
\end{equation}
where $\beta_2$ is given by~\eqref{beta_2}. This agrees
with Malik and Wands equations (6.27)
and (6.28) apart from a differing sign in (6.27).\footnote{We find that after setting
$\xi_1^i=0$ (6.27) should read $\chi_{{\mathrm v}i}= 2\alpha_1(v_{1i}' - {\cal H}v_{1i})$.}

Finally, for ${}^{(2)}\!{\bdelta}$ the un-scaled form of our
equation~\eqref{trans_delta2} in Malik and Wands
notation is as follows:
\begin{equation}
{\tilde\rho}_2 = \rho_2 + \rho_0' \,\alpha_2 + \alpha_1(\rho_0''\, \alpha_1 + \rho_0'\,\alpha_1' +2\rho_1'), \label{brho_MW}
\end{equation}
which agrees with Malik and Wands equation (6.20) after setting $\xi_1^i=0$.

\section{Spatial differential operators~\label{spat.diff.op}}

In this appendix we introduce the spatial
differential operators that are used in this paper. First, we require
the spatial Laplacian and the trace-free symmetric second order derivative:
\begin{equation} \label{2order_D}
{\bf D}^2 := \gamma^{ij}{\bf D}_i{\bf D}_j, \qquad
{\bf D}_{ij} := {\bf D}_{(i}{\bf D}_{j)} - \sfrac13 \gamma_{ij}{\bf D}^2.
\end{equation}
We use these to define the \emph{scalar mode extraction
operators} (see~\cite{uggwai13b}, equations (85)):\footnote{If the background is  not flat,
${\cal S}^{ij} = \sfrac32 {\bf D}^{-2}({\bf D}^2 +K)^{-1}{\bf D}^{ij}$.}
\begin{subequations}
\begin{equation}\label{modeextractop}
{\cal S}^{i} = {\bf D}^{-2}{\bf D}^i ,\qquad
{\cal S}^{ij} = \sfrac32 ({\bf D}^{-2})^2{\bf D}^{ij},
\end{equation}
where ${\bf D}^{-2}$ is the inverse spatial Laplacian. Note that
${\cal S}^{i}$ is the inverse operator of ${\bf D}_i$ and
${\cal S}^{ij}$ is the inverse operator of ${\bf D}_{ij}$:
\be {\cal S}^{i}{\bf D}_i C=C, \qquad   {\cal S}^{ij}{\bf D}_{ij}C=C. \ee
\end{subequations}
 We can now
define the following spatial differential operators:
\begin{subequations} \label{D0,2}
\begin{align}
({\bf D} C)^2 &:=\gamma^{ij}({\bf D}_iC)({\bf D}_jC), \label{QD1}  \\
{\mathbb D}_0(C) &:= {\cal S}^{ij}({\bf D}_iC)({\bf D}_jC), \label{QD2} \\
{\mathbb D}_2(C) &:= \sfrac13\left(
{\bf D}^2{\mathbb D}_0(C) - ({\bf D}C)^2 \right), \label{QD3}
\end{align}
\end{subequations}
which act on an arbitrary function $C$.

The operator $({\bf D} C)^2 $ is familiar, being the square of the magnitude
of the gradient ${\bf D}_i C$, whereas the other two are less so.
The expression ${\mathbb D}_0(C)$ can be viewed as the scalar mode
of the rank two tensor $({\bf D}_iC)({\bf D}_jC)$ while ${\mathbb D}_2(C)$
is defined  by taking the Laplacian of ${\mathbb D}_0(C)$.
We mention one property of these operators that suggests their physical role.
When taking limits such as the long wavelength limit
and the Newtonian limit it is essential to count
how quantities change under a scaling of the
spatial coordinates. More precisely, if some expression
$L({\bf D}_i)$ involving ${\bf D}_i$ scales as
$L(\lambda{\bf D}_i)= {\lambda}^p L({\bf D}_i)$ under a rescaling of spatial coordinates
$x^i\rightarrow {\lambda}^{-1}x^i$,
we say that $L({\bf D}_i)$ has
weight $p$ in ${\bf D}_i$.\footnote{Note that ${\bf D}^2$ and ${\bf D}_{ij}$
both have weight $2$ while in contrast ${\cal S}^{i}$
and ${\cal S}^{ij}$ have weights $-1$ and $-2$,
respectively.}
It follows from equations~\eqref{modeextractop}
and~\eqref{D0,2} that
${\mathbb D}_0(C)$ has weight $0$ while ${\mathbb D}_2(C)$ has
weight $2$ in ${\bf D}_i$.
We thus expect that ${\mathbb D}_0(C)$ will be dominant
and ${\mathbb D}_2(C)$ will be negligible in the
long wavelength limit, while the reverse will be true in the Newtonian
limit.

In this paper the operators ${\mathbb D}_0$
and ${\mathbb D}_2$ serve to simplify the quadratic source
terms in the gauge change formulas at second order. They
play a similar role in other source terms, for example the source terms
of the perturbed Einstein tensor, and hence in the solutions of
the perturbed Einstein equations at second order.
The fact that these operators occur frequently motivates our choice of notation:
the symbol ${\mathbb D}$ suggests a spatial differential
operator acting on an arbitrary function $C$, while the subscript ${}_0$ or ${}_2$
indicates the weight in ${\bf D}_i$. The use of ${\bf D}_i$
in general, rather than ${}_{,i}$ also helps
to clarify the structure of the spatial derivative terms.

It turns out that the operators ${\mathbb D}_0$
and ${\mathbb D}_2$, as defined in~\eqref{D0,2}, are related to two quantities,
denoted $\Psi_0$ and $\Theta_0$, that have been used
in the literature on second order perturbations since 1997.
These quantities were defined in the case of a flat background
by Mollerach and Matarrese (1997)~\cite{molmat97} as
follows:\footnote{See equation(3.7) for $\Psi_0$ and
equation (3.11) for $\Theta_0$ in~\cite{molmat97}.
See also Materrese \emph{et al} (1998)~\cite{matetal98},
equations (4.36) and (6.6). }
\begin{subequations} \label{psi_0,theta_0}
\begin{align}
\Psi_0 &:=\sfrac12 {\bf D}^{-2}\left( {\bf D}^i{\bf D}^j C\, {\bf D}_i{\bf D}_j C-({\bf D}^2C)^2 \right),  \label{D7a}  \\
\Theta_0 &:={\bf D}^{-2}\left(\Psi_0 -\sfrac13 ({\bf D}C)^2 \right), \label{D8a}
\end{align}
\end{subequations}
where $C$ is a  function that determines the spatial dependence of the
perturbations.
The explicit relations are simple, namely
\begin{equation}
\Theta_0 = -\sfrac13 {\mathbb D}_0(C), \qquad  \Psi_0 = - {\mathbb D}_2(C), \label{D_0,2.relation}
\end{equation}
but their derivation requires the use of some complicated identities
satisfied by ${\bf D}_i$, as follows.
Expand the second derivatives on the left sides to get
\begin{subequations}
\begin{align}
{\bf D}^i{\bf D}^j({\bf D}_i C{\bf D}_j C)&= 2({\bf D}_i C ){\bf D}^i ({\bf D}^2 C) +
({\bf D}^i{\bf D}^jC)( {\bf D}_i{\bf D}_j C) + ({\bf D}^2 C)^2,  \\
{\bf D}^2 ({\bf D}C)^2&=2({\bf D}_i C ){\bf D}^i ({\bf D}^2 C) +
2({\bf D}^i{\bf D}^jC)( {\bf D}_i{\bf D}_j C).
\end{align}
\end{subequations}
Next take the difference and replace ${\bf D}^i{\bf D}^j$ by
${\bf D}^{ij}$ on the left side to obtain:
\be {\bf D}^{ij}({\bf D}_i C{\bf D}_j C)- \sfrac23{\bf D}^2 ({\bf D}C)^2=
-({\bf D}^i{\bf D}^jC)( {\bf D}_i{\bf D}_j C)+ ({\bf D}^2 C)^2, \label{key_identity} \ee
which is the key identity.
The desired relations~\eqref{D_0,2.relation} follow immediately
from~\eqref{key_identity} on using the
definitions~\eqref{D0,2} and~\eqref{psi_0,theta_0}.

Although $\Theta_0$ and $\Psi_0$ were first introduced to describe second order perturbations of the Einstein-de Sitter universe, since 2005 they
have also been used to describe perturbed $\Lambda$CDM universes.
 See, for example, Bartolo \emph{et al}
(2005)~\cite{baretal05} following equation (9), Tomita (2005)~\cite{tom05},
equations (2.11),
Villa and Rampf (2016)~\cite{vilram16}, equations (5.21) and (5.38), and
Tram \emph{et al} (2016)~\cite{traetal16}, equations (D.11) and (D.12).
In these references one sees that $\Theta_0$ (\emph{i.e.} $\!{\mathbb D}_0(C)$)
contributes to perturbations on super-horizon scales, while ${\bf D}^2\Psi_0$
(\emph{i.e.} $\!{\bf D}^2{\mathbb D}_2(C)$) contributes to the
Newtonian part of the second order density perturbation.
As mentioned earlier, this physical
interpretation\footnote{In a suggestion of physical interpretation Villa and Rampf (2016)~\cite{vilram16} (see
page 15 following equation (5.39)) refer to $\Theta_0$
as the "GR kernel" and to $\Psi_0$ as the ``Newtonian kernel.''}
is a consequence of the fact that ${\mathbb D}_0(C)$
is of weight zero in ${\bf D}_i$ while ${\mathbb D}_2(C)$ is of weight two and
${\bf D}^2{\mathbb D}_2(C)$ is of weight four.
The important point, however, is that ${\mathbb D}_0(C)$
 and ${\mathbb D}_2(C)$ are not restricted in use
 to the $\Lambda$CDM universes: they arise in the general change of gauge
formulas in this paper, in the  source terms of
the second order perturbations of the Einstein tensor,
and ${\mathbb D}_0(C)$ contributes to perturbations
on super-horizon scales in general. For example,
${\mathbb D}_0(\psi_{\mathrm p})$ contributes to
the CMB anisotropy at second order on large scales.\footnote
{See for example Bartolo \emph{et al} (2010)~\cite{baretal10}, equation (53).}

\end{appendix}

\bibliographystyle{plain}
\bibliography{../Bibtex/cos_pert_papers}

\begin{thebibliography}{10}

\bibitem{baretal04a}
N.~Bartolo, E.~Komatsu, S.~Matarrese, and A.~Riotto.
\newblock Non-{G}aussianity from inflation: theory and observations.
\newblock {\em Physics Reports}, {\bf 402}:103--266, 2004.

\bibitem{baretal05}
N.~Bartolo, S.~Matarrese, and A.~Riotto.
\newblock Signatures of primordial non-{G}aussianity in the large-scale
  structure of the universe.
\newblock {\em JCAP}, {\bf 0510}:010, 2005.

\bibitem{baretal10}
N.~Bartolo, S.~Matarrese, and A.~Riotto.
\newblock Non-{G}aussianity and the cosmic microwave background anisotropies.
\newblock {\em Advances in Astronomy}, {\bf 2010}:157079, 2010.

\bibitem{bruetal97}
M.~Bruni, S.~Matarrese, S.~Mollerach, and S.~Sonego.
\newblock Perturbations of spacetime: Gauge transformations and gauge
  invariance at second order and beyond.
\newblock {\em Class. Quantum Grav.}, {\bf 14}:2585, 1997.

\bibitem{carmal15}
P.~Carrilho and K.~A. Malik.
\newblock Vector and tensor contributions to the curvature perturbation at
  second order.
\newblock {\em JCAP}, {\bf 02}:021, 2015.

\bibitem{chretal11b}
A.~J. Christopherson, K.~A. Malik, D.~R. Matravers, and K.~Nakamura.
\newblock Comparing different formulations of nonlinear perturbation theory.
\newblock {\em Class. Quantum Grav.}, {\bf 28}:225024, 2011.

\bibitem{chretal15}
A.~J. Christopherson, E.~Nalson, and K.~A. Malik.
\newblock A short note on the curvature perturbation at second order.
\newblock {\em Class. Quantum Grav.}, {\bf 32}:075005, 2015.

\bibitem{diaetal15}
M.~Dias, J.Elliston, J.~Frazer, D.~Mulryne, and D.~Seery.
\newblock The curvature perturbation at second order.
\newblock {\em JCAP}, {\bf 02}:040, 2015.

\bibitem{grebru17}
H.~A. Gressel and M.~Bruni.
\newblock fnl - gnl mixing in the matter density field at higher orders.
\newblock {\em JCAP}, (06):016, 2018.

\bibitem{hidetal13}
J.~C. Hidalgo, A.~J. Christopherson, and K.~A. Malik.
\newblock The {P}oisson equation at second order in relativistic cosmology.
\newblock {\em JCAP}, {\bf 08}:026, 2013.

\bibitem{hwaetal17}
J-C. Hwang, D.~Jeong, and H.~Noh.
\newblock Gauge dependence of gravitational waves generated from scalar
  perturbations.
\newblock {\em Astrophysical Journal}, {\bf 842}:46, 2017.

\bibitem{kodsas84}
H.~Kodama and M.~Sasaki.
\newblock Cosmological perturbation theory.
\newblock {\em Prog. Theoret. Phys. Suppl.}, {\bf 78}:1--166, 1984.

\bibitem{lidlyt00}
A.~R. Liddle and D.~H. Lyth.
\newblock {\em Cosmological inflation and large-scale structure}.
\newblock Cambridge University Press, 2000.

\bibitem{lytrod05}
D.~H. Lyth and Y.~Rodriguez.
\newblock Non-gaussianity from the second-order cosmological perturbation.
\newblock {\em Phys. Rev. D}, {\bf 71}:123508, 2005.

\bibitem{mal05}
K.~A. Malik.
\newblock Gauge-invariant perturbations at second order - multiple scalar
  fields on large scales.
\newblock {\em JCAP}, {\bf 0511}:005, 2005.

\bibitem{malwan09}
K.~A. Malik and D.~Wands.
\newblock Cosmological perturbations.
\newblock {\em Physics Reports}, {\bf 475}:1--51, 2009.

\bibitem{marrin06}
J.~Martin and C.Ringeval.
\newblock Inflation after wmap3: confronting the slow-roll and exact power
  spectra with cmb data.
\newblock {\em JCAP}, {\bf 08}:009, 2006.

\bibitem{matetal98}
S.~Matarrese, S.~Mollerach, and M.~Bruni.
\newblock Relativistic second-order perturbations of the {E}instein-de {S}itter
  universe.
\newblock {\em Phys.\ Rev.\ D}, {\bf 58}:043504, 1998.

\bibitem{molmat97}
S.~Mollerach and S.~Matarrese.
\newblock Cosmic microwave background anisotropies from second order
  gravitational perturbations.
\newblock {\em Phys. Rev. D}, {\bf 56}:4494, 1997.

\bibitem{nak07}
K.~Nakamura.
\newblock Second order gauge invariant cosmological perturbation theory.
\newblock {\em Prog. Theoret. Phys.}, {\bf 117}:17, 2007.

\bibitem{nohhwa04}
H.~Noh and J-C. Hwang.
\newblock Second order perturbations of the {F}riedmann world model.
\newblock {\em Phys. Rev. D}, {\bf 69}:104011, 2004.

\bibitem{salbon90}
D.~S. Salopek and J.R. Bond.
\newblock Nonlinear evolution of long-wavelength metric fluctuations in
  inflationary models.
\newblock {\em Phys. Rev. D}, {\bf 42}:3936, 1990.

\bibitem{tom05}
K.~Tomita.
\newblock Relativistic second-order perturbations of nonzero-$\lambda$ flat
  cosmological models and {C}{M}{B} anisotropies.
\newblock {\em Phys. Rev. D}, {\bf 71}:083504, 2005.

\bibitem{traetal16}
T.~Tram, C.~Fidler, R.~Crittenden, K.~Koyama, G.~W. Pettinari, and D.~Wands.
\newblock The intrinsic matter bispectrum in {$\Lambda$CDM}.
\newblock {\em JCAP}, {\bf 05}:058, 2016.

\bibitem{uggwai13b}
C.~Uggla and J.~Wainwright.
\newblock Asymptotic analysis of perturbed dust cosmologies to second order.
\newblock {\em Gen. Rel. Grav.}, {\bf 45}:1467, 2013.

\bibitem{uggwai14a}
C.~Uggla and J.~Wainwright.
\newblock Second order density perturbations for dust cosmologies.
\newblock {\em Phys. Rev. D}, {\bf 90}:043511, 2014.

\bibitem{uggwai18}
C.~Uggla and J.~Wainwright.
\newblock Second order cosmological perturbations: dynamics.
\newblock {\em Phys. Rev. D}, {\bf 98}:103534, 2018.

\bibitem{uggwai19b}
C.~Uggla and J.~Wainwright.
\newblock Second order cosmological perturbations: conserved quantities and
  explicit solutions at large scales.
\newblock {\em Preprint}, 2019.

\bibitem{uggwai19c}
C.~Uggla and J.~Wainwright.
\newblock Single field inflationary universes: the general solution at large
  scale for second order perturbations.
\newblock {\em Preprint}, 2019.

\bibitem{ver05}
F.~Vernizzi.
\newblock On the conservation of second-order cosmological perturbations in a
  scalar field dominated universe.
\newblock {\em Phys. Rev. D}, {\bf 71}:06130, 2005.

\bibitem{vilram16}
E.~Villa and C.~Rampf.
\newblock Relativisitic perturbations in {$\Lambda$}{CDM}: {E}ulerian and
  {L}agrangian approaches.
\newblock {\em JCAP}, {\bf 01}:030, 2016.

\end{thebibliography}

\end{document}